%Paper: hep-ph/9501320
%From: hhaber@surya11.cern.ch (Howard Haber)
%Date: Tue, 17 Jan 1995 15:59:21 +0100 (MET)

\def\SCIPP{SCIPP}%%%
\def\quarter{{1\over4}}%%%
\input phyzzx.tex
%Uses the PHYZZX macropackage
\hoffset=1cm
\overfullrule0pt
\hsize=35.5pc \vsize=51pc
\normalparskip=0pt
\parindent=36pt
\itemsize=36pt
\def\ifmath#1{\relax\ifmmode #1\else $#1$\fi}
\def\cb{c_{\beta}}

\def\hl{h^0}
\def\ha{A^0}
\def\hh{H^0}
\def\hm{H^-}
\def\hp{H^+}
\def\hpm{H^{\pm}}
\def\mha{m_{\ha}}
\def\mhh{m_{\hh}}
\def\mhl{m_{\hl}}
\def\mhpm{m_{\hpm}}
\def\mt{m_t}
\def\mw{m_W}
\def\wm{W^-}
\def\mz{m\ls Z}
\let\rta=\to
\def\sb{s_{\beta}}
\def\tanb{\tan\beta}

\def\mstop{M_{\tilde t}}
\def\mlsq{m\ls{L}^2}
\def\mtsq{m\ls{T}^2}
\def\mdsq{m\ls{D}^2}
\def\mssq{m\ls{S}^2}
\def\calm{{\cal M}}
\def\calv{{\cal V}}
\def\crrr{\cr\noalign{\vskip8pt}}
\def\eighth{\ifmath{{\textstyle{1 \over 8}}}}
\def\GENITEM#1;#2{\par\vskip6pt \hangafter=0 \hangindent=#1
   \Textindent{$ #2$ }\ignorespaces}

\def\unlock{\catcode`@=11} % This allows us to modify PLAIN macros.
\def\lock{\catcode`@=12} % at signs are no longer letters
\unlock
%%%%  TO LOWER A SUBSCRIPT
\def\ls#1{_{\lower1.5pt\hbox{$\scriptstyle #1$}}}
\def\refitem#1{\r@fitem{#1.}}
\def\refout{\par\penalty-400\vskip\chapterskip
   \spacecheck\referenceminspace
   \ifreferenceopen \Closeout\referencewrite \referenceopenfalse \fi
   \line{\hfil \bf References\hfil}\vskip\headskip
   \input \jobname.refs
   }
\chapterminspace=3pc
\def\chapter#1{\par \penalty-300 \vskip2pc
   \spacecheck\chapterminspace
   \chapterreset \leftline{\bf \chapterlabel.~~#1}
   \nobreak\vskip\headskip \penalty 30000
   {\pr@tect\wlog{\string\chapter\space \chapterlabel}} }
\def\section#1{\par \ifnum\lastpenalty=30000\else
   \penalty-200\vskip\sectionskip \spacecheck\sectionminspace\fi
   \gl@bal\advance\sectionnumber by 1
   {\pr@tect
   \xdef\sectionlabel{\chapterlabel.%
       \the\sectionstyle{\the\sectionnumber}}%
   \wlog{\string\section\space \sectionlabel}}%
   \noindent {\it\sectionlabel.~~#1}\par
   \nobreak\vskip\headskip \penalty 30000 }\lock
%
% R.G. Leigh 20 July 1993
%
%\def\treetitle{
%\let\picnaturalsize=N
%\ifx\twocolin Y
%       \def\picsize{1.2in}
%       \else\def\picsize{1.75in}\fi
%\def\picfilename{scipp_tree.eps}
%\centerline{} \vskip-3pc
%\ifx\nopictures Y
%       \hfill\vbox{\hbox{\the\Pubnum}\hbox{\the\date}}
%       \else{\ifx\epsfloaded Y
%               \else\input epsf \fi
%       \let\epsfloaded=Y
%       {\line{\hbox{\ifx\picnaturalsize N\epsfysize \picsize\fi
%       {\epsfbox{\picfilename}}}\hfill\vbox{
%       \hbox{\the\Pubnum}\hbox{\the\date}
%       \ifx\twocolin Y
%               \vskip1.3in
%               \else\vskip1in\fi
%       }}}}\fi
%}
%

%%%%%%%%%%%%%%%%%%%%%%%%%%%%%%%%%%%%%%%%%%%%%%%%%%%%%%%%%%%%%%%%%%%%
\frontpagetrue
\Pubnum={SCIPP 94/39}
\date={December 1994}
%\titlepage
\centerline{} \vskip-3pc
%\treetitle
\singlespace
\vbox to .5cm{}
\centerline{{\fourteenbf  Non-Minimal Higgs Sectors: The Decoupling Limit}}
\vskip3pt
\centerline{{\fourteenbf and its Phenomenological Implications}
\foot{Work supported in part by the U.S.~Department of Energy.}}
\vskip1cm
\centerline{\caps Howard E. Haber}
\vskip .1in
\SCIPP
\vskip1cm
\vbox{ \narrower
\centerline{\bf Abstract}
\vskip6pt
In models with a non-minimal Higgs sector,  a decoupling limit can be defined.
In this limit, the masses of all the physical Higgs states are large
 (compared to the scale of
electroweak symmetry breaking) except for one neutral CP-even Higgs scalar,
whose properties are indistinguishable from the Higgs boson of the minimal
Standard Model.   The decoupling limit of the most general CP-conserving
two-Higgs doublet model is formulated.   Detection of evidence for a
non-minimal Higgs
sector at future colliders in the decoupling limit may present a
formidable challenge for future Higgs searches.
}
\vfill
\centerline{Invited Talk presented at the}
\centerline{Workshop on Electroweak Symmetry Breaking, E\"otv\"os University}
\centerline{Budapest, Hungary,  11--13 July 1994, }
\centerline{and at the}
\centerline{Workshop on Physics from the Planck Scale to the Electroweak
 Scale,}
\centerline{University of Warsaw, Poland, 21--24 September, 1994.}
\vfill
\endpage
\bigskip
\chapter{Introduction}
\medskip

With the recent ``discovery'' of the top quark, the only missing
piece of the Standard Model is the Higgs boson.  The Standard Model
posits the existence of one complex weak doublet (with
hypercharge $Y=1$) of elementary scalar (Higgs) fields.  The dynamics
of these fields is assumed to trigger electroweak symmetry breaking.
Three Goldstone bosons are absorbed by the $W^\pm$ and $Z$ leaving
one CP-even neutral Higgs scalar to be discovered.

\REF\hhg{J.F. Gunion, H.E. Haber, G.L. Kane and S. Dawson,
{\it The Higgs Hunter's Guide} (Addison-Wesley, Redwood City, CA, 1990).}
\REF\miransky{V.A. Miransky, {\it Dynamical Symmetry Breaking in
Quantum Field Theories} (World Scientific, Singapore, 1993).}
Despite the enormous success of the Standard Model, the existing
data sheds little light on the spectrum and dynamics of
the electroweak symmetry breaking sector.  The scalar spectrum could
be minimal (as in the Standard Model) or non-minimal, with a rich
spectrum of scalar states (including perhaps bound states of higher spin).
The dynamics could involve either weakly interacting or strongly
interacting forces.  Many examples displaying each of the above
features have been studied in the literature; comprehensive reviews
can be found is refs.~\hhg\ and \miransky.   In this paper, I will
assume that electroweak symmetry breaking is a result of the dynamics
of a weakly-coupled Higgs sector.  Such a theory may be technically
natural if embedded in a theory of low-energy supersymmetry.
However, the results obtained in this work do not necessarily
require the existence of low-energy supersymmetry.

\REF\cpr{P.H. Chankowski, S. Pokorski and J. Rosiek, {\sl Phys. Lett.}
{\bf B281} (1992) 100.}
\REF\gkane{G.L. Kane, in {\it Perspectives on Higgs Physics}, edited by
G.L. Kane (World Scientific, Singapore, 1993) pp.~223--228.}
In this paper, I shall pose the following questions.  Assume that a
scalar state (\ie, a candidate for the Higgs boson) is discovered in
a future collider experiment.  What are the expectations for its
properties?  Will it resemble the Higgs boson of the minimal Standard
Model or will it possess some distinguishing trait?  If the
properties of this scalar state are difficult to distinguish from the
Standard Model Higgs boson, what are the requirements of future
collider experiments for detecting the existence (or non-existence)
of a non-minimal Higgs sector?   Some of these questions have also been
addressed by other authors; see \eg, refs.~\cpr\ and \gkane.

These questions would be moot if the experiment that first
discovers the lightest Higgs scalar also discovers a second scalar state.  For
example, in a theory with a light CP-even scalar ($\hl$) and a light
CP-odd scalar ($\ha$), both states would be produced in
$e^+e^-$ collisions via $s$-channel $Z$-exchange
($e^+e^-\rta Z\rta\hl\ha$) if kinematically allowed.
The simultaneous discovery of $\hl$ and $\ha$ would clearly
indicate the existence of a non-minimal Higgs sector.  In this paper,
I shall not consider such a scenario.  As we shall see,
the case where only one light scalar state is initially discovered may present
a formidable challenge to unraveling the underlying
structure of the electroweak symmetry breaking sector.

For simplicity, I focus in this paper on the (CP-conserving)
two-Higgs doublet model.  In this model, the physical scalar states
consist of a charged Higgs pair ($\hpm$), two CP-even scalars ($\hl$
and $\hh$, with $\mhl\le\mhh$) and one CP-odd scalar ($\ha$).
The ultimate conclusions of this paper will
survive in models with more complicated scalar sectors.
Following the discussion above, the working hypothesis of this
paper is that $\hl$, assumed to be the lightest scalar state,
 will be the first Higgs boson to be discovered.
%\foot{One can consider other scenarios where the lightest
%scalar is not $\hl$.  I will briefly address this possibility in
%section ***.}
Moreover,
the mass gap between $\hl$ and the heavier scalars is assumed to be
sufficiently large so that the initial experiments which can detect $\hl$
will not have sufficient energy and luminosity to initially discover
any of the heavier scalar states.

\REF\pdg{L. Montanet \etal\ [Particle Data Group] {\sl Phys. Rev.}
{\bf D50} (1994) 1173.}
\REF\sopczak{A. Sopczak, {\sl Int. J. Mod. Phys.} {\bf A9}
(1994) 1747.}
\REF\mrenna{A. Stange, W. Marciano and S. Willenbrock, {\sl Phys. Rev.}
{\bf D49} (1994) 1354; {\bf D50} (1994) 4491; S. Mrenna and G.L. Kane,
CALT-68-1938 (1994) [hep-ph 9406337]; J.F. Gunion, UCD-94-24
(1994) [hep-ph 9406405].}
\REF\heavyhiggs{D. Froidevaux, Z. Kunszt and J. Stirling \etal, in
{\it Proceedings of the Large Hadron Collider Workshop},
Aachen 1990, CERN Repoprt 90-10 (1990).}
\REF\imh{S. Dawson, in {\it Perspectives on Higgs Physics}, edited
by G.L. Kane (World Scientific, Singapore, 1993) pp.~129--155; Z. Kunszt,
{\it ibid.}~pp.~156--178.}
\REF\bargeretal{V. Barger, K. Cheung, A. Djouadi, B.A. Kniehl, and
P. Zerwas, {\sl Phys. Rev.} {\bf D49} (1994) 79.}
\REF\ghphoton{J.F. Gunion and H.E. Haber,
{\sl Phys. Rev.} {\bf D48} (1993) 5109.}
\REF\caldwell{D.L. Borden, D.A. Bauer and D.O. Caldwell,
{\sl Phys. Rev.} {\bf D48} (1993) 4018.}
How will $\hl$ be discovered and where?  Present LEP bounds\refmark\pdg\
imply
that $\mhl\gsim 60$~GeV.  This bound will be improved by
LEP-II,\refmark\sopczak\
which will be sensitive to Higgs masses up to roughly $\sqrt{s}-\mz-
10$~GeV.  The LEP search is based on $e^+e^-\rta Z\rta Z\hl$ where
one of the two $Z$'s is on-shell and the other is off-shell.   At hadron
colliders, an upgraded Tevatron with an integrated luminosity of
10~fb$^{-1}$ can begin to explore the intermediate-mass Higgs
regime\refmark\mrenna\
($80\lsim\mhl\lsim 130$~GeV).  The Higgs search at the LHC will
significantly extend the Higgs search to higher masses\refmark\heavyhiggs\
(although the intermediate mass regime still presents some significant
difficulties for the LHC
detector collaborations).  The dominant mechanism for Higgs
production at hadron colliders is via $gg$-fusion through a top-quark
loop.  If $\mhl>2\mz$, the ``gold-plated'' detection mode is $\hl\rta
ZZ$; each $Z$ subsequently decays leptonically, $Z\rta\ell^+\ell^-$
(for $\mhl\gsim 130$~GeV, $\hl\rta ZZ^\ast$, where $Z^\ast$ is
off-shell, provides a viable signature).  Other decay modes are required
in the case of the intermediate mass Higgs
(for recent reviews, see ref.~\imh).
At a future $e^+e^-$ linear collider (NLC), the Higgs mass
reach of LEP-II will be extended.\refmark\bargeretal\  In addition,
with increasing $\sqrt{s}$, Higgs boson production via
$W^+W^ -$ fusion begins to be the dominant production process.
Finally, one novel possibility is to run the NLC in a $\gamma\gamma$-%
collider mode.  In this mode, Higgs production via $\gamma\gamma$-%
fusion (which is typically dominated by a $\wp\wm$ and/or a $t\bar t$
loop) may be detectable, depending on the particular Higgs final
state decay.\refmark{\ghphoton, \caldwell}\
Note that almost all of the Higgs search techniques outlined
above involve the $\hl ZZ$ (and in some cases the
$\hl\wp\wm$) vertex.   In a few cases, it is the $\hl t\bar t$ vertex
(and possibly the $\hl b\bar b$ vertex) that plays the key role.
These observations are relevant for the phenomenological
considerations of this paper.

In section 2, I review the Higgs sector of the minimal supersymmetric
extension of the Standard Model (MSSM).  In this context, I discuss
why one might expect that $\hl$ is the lightest scalar whose
properties are nearly identical to that of the Standard Model Higgs
boson.  In section 3, I place the results of section 2 in a more
general context.  I define the ``decoupling limit'' of the general
two-Higgs doublet model; in this limit, $\hl$ is indistinguishable from
the Standard Model Higgs boson.  In section 4, I discuss the
phenomenological challenges of the decoupling limit for the Higgs search
at future
colliders.  After briefly mentioning the prospects for non-minimal
Higgs detection at the LHC, I consider in more detail the prospects
for the discovery of the non-minimal Higgs sector at the NLC.
Conclusions are presented in section 5.
\bigskip

\chapter{The Higgs Sector of the MSSM---A Brief Review}
\medskip

\REF\hhgsusy{J.F. Gunion and H.E. Haber, {\sl Nucl. Phys.} {\bf B272}
(1986) 1; {\bf B278} (1986) 449 [E: {\bf B402} (1993) 567].}
%A comprehensive review of the tree-level Higgs sector
%of the MSSM can be found in ref.~\hhg.
In this section, I provide a very brief review of the Higgs sector of
the minimal supersymmetric extension of the Standard Model
(MSSM).\refmark\hhgsusy\
%We shall see that over nearly the entire range of supersymmetric
%parameter space, $\hl$ is the lightest Higgs state whose couplings to
%Standard Model particles (quarks, leptons and gauge bosons) are
%nearly identical to the Higgs boson of the minimal Standard Model.First,
let us consider the MSSM Higgs potential at tree-level
which depends on two complex doublet scalar fields $H_1$ and $H_2$
of hypercharge $-1$ and $+1$, respectively :
$$\eqalign{%
  \calv = \ &m^2_{11} |H_1|^2 + m^2_{22} |H_2|^2
                 - m^2_{12} (\epsilon_{ij} H^i_1 H^j_2 + {\rm h.c.})\crr
                 &+\eighth(g^2+g'^2) \left(|H_1|^2-|H_2|^2\right)^2
                    +\half g^2|H^*_1H_2|^2\,,\cr
}\eqn\mssmhiggspot$$
where $m^2_{ii}\equiv|\mu|^2+m^2_i\quad (i=1,2)$.
The parameters $m_1^2$, $m_2^2$ and $m_{12}^2$ are
soft-supersymmetry-breaking parameters, $\mu$ is the
Higgs superfield mass parameter, and $g$ and $g^\prime$ are the
SU(2)$\times$U(1) gauge couplings.

The three parameters $m^2_{11},\ m^2_{22}$ and $m^2_{12}$ of the
Higgs potential can re-expressed in terms of the two Higgs vacuum
expectation values, $\VEV{H_i^0}\equiv v_i/\sqrt{2}$
and one physical Higgs
mass.  One is free to choose the phases of the Higgs fields such
that $v_1$ and $v_2$ are positive.   Then, $m_{12}^2$ must be positive,
in which case it follows from eq.~\mssmhiggspot\ that
$$
  \mha^2 = {m^2_{12}\over \sin\beta \cos\beta}\,.\eqn\monetwoeq
$$
%It is convenient to choose $m_A$ as one of the physical input
%parameters.
Note that $\mw^2=\quarter
g^2(v_1^2+v_2^2)$, which fixes the magnitude $v_1^2+v_2^2=(246~{\rm
GeV})^2$.  This leaves two parameters which determine all the Higgs sector
masses and couplings: $\mha$ and $\tan\beta\equiv v_2/v_1$.
The charged Higgs mass is given by
$$
  m^2_{H^\pm} = m^2_W +\mha^2\,.\eqn\chhiggs
$$
The neutral CP-even Higgs bosons, $H^0$ and $h^0$, are obtained by
diagonalizing a $2\times 2$ mass matrix.
%which in the $H_1$--$H_2$
%basis is given by
%$$
%  {\cal M}^2 = \pmatrix{%
%   \mha^2 s^2_\beta + m^2_Z c^2_\beta &-(\mha^2+m^2_Z)s_\beta c_\beta
%   \cr
%  -(\mha^2+m^2_Z)s_\beta c_\beta &\mha^2 c^2_\beta+m^2_Zs^2_\beta \cr }
%\, .\eqn\twotimestwo
%$$
The eigenstates are
$$\eqalign{H^0 &= (\sqrt 2\,{\rm Re}\,H^0_1-v_1) \cos\alpha +
(\sqrt2\,{\rm Re}\,H^0_2-v_2)
              \sin\alpha   \cr
  h^0 &= -(\sqrt 2\,{\rm Re}\,H^0_1-v_1)\sin\alpha + (\sqrt2\,{\rm Re}
\,H^0_2-v_2)
             \cos\alpha \cr
}\eqn\cpevenhiggs$$
which defines the CP-even Higgs mixing angle $\alpha$.
%In the MSSM,
%with the phase choices as specified above, $-\pi/2\leq\alpha\leq 0$.
%where the mixing angle $\alpha$ is given by
%$$
%  \cos 2\alpha = -\cos2\beta \left( {\mha^2-m^2_Z\over
%                   m^2_{H^0}-m^2_{h^0}}\right),
%\qquad\quad \sin2\alpha = -\sin2\beta \left({m^2_{H^0}+m^2_{h^0} \over
%                   m^2_{H^0}-m^2_{h^0}}\right)\,.\eqn\alphadef
%$$
The corresponding CP-even Higgs mass eigenvalues are
$$
  m^2_{H^0,\,h^0} = \half \left(\mha^2+m^2_Z \pm
       \sqrt{(\mha^2+m^2_Z)^2 - 4m^2_Z \mha^2 \cos^2 2\beta}\right)\,,
\eqn\cpevenhiggsmass
$$
where by definition, $\mhl\leq\mhh$.  Explicit formulae for $\alpha$ can also
be derived.  Here, I shall note one particularly useful relation
$$\cos^2(\beta-\alpha)={\mhl^2(\mz^2-\mhl^2)\over
\mha^2(\mhh^2-\mhl^2)}\,.\eqn\cbmasq$$

Consider the limit where $\mha\gg\mz$.  Then, from the above
formulae,
$$\eqalign{
\mhl^2\simeq\ &\mz^2\cos^2 2\beta\,,\cr
\mhh^2\simeq\ &\mha^2+\mz^2\sin^2 2\beta\,,\cr
\mhpm^2=\ & \mha^2+\mw^2\,,\cr
\cos^2(\beta-\alpha)\simeq\ &{\mz^4\sin^2 4\beta\over 4\mha^4}\,.\cr}
\eqn\largema$$Two consequences are immediately apparent.
First, $\mha\simeq\mhh
\simeq\mhpm$, up to corrections of ${\cal O}(\mz^2/\mha)$.  Second,
$\cos(\beta-\alpha)=0$ up to corrections of ${\cal O}(\mz^2/\mha^2)$.

\REF\impact{For a review of the influence of radiative corrections on the
MSSM Higgs sector, see H.E. Haber, in {\it Perspectives on Higgs
Physics}, edited by G.L. Kane (World Scientific, Singapore, 1993) pp.~79--128.}
\REF\radcorrs{See, \eg, H.E. Haber and R. Hempfling, {\sl Phys. Rev.}
{\bf D48} (1993) 4280.}
It is known that one-loop radiative corrections have a
significant impact on the MSSM Higgs sector.\refmark\impact\   Nevertheless,
the conclusions just stated are not modified.  For example, in the
limit where $\mha\gg\mz$ and $\mz\ll\mt\ll\mstop$, the one-loop
radiatively corrected Higgs squared masses are\refmark\radcorrs\
$$\eqalign{
\mhl^2\simeq\ &\mz^2\cos^2 2\beta+{3g^2\over 8\pi^2\mw^2}
\left[\mt^4+\half\mt^2\mz^2\cos 2\beta\right]
\ln\left({\mstop^2\over\mt^2}\right)\,,\crr
\mhh^2\simeq\ &\mha^2+\mz^2\sin^2 2\beta+{3g^2\cos^2\beta\over 8\pi^2\mw^2}
\left[{\mt^4\over\sin^2\beta}-\mt^2\mz^2\right]
\ln\left({\mstop^2\over\mt^2}\right)\,,\crr
\mhpm^2\simeq\ &\mha^2+\mw^2+{3g^2\over 32\pi^2}\left[{2\mt^2 m_b^2\over
\mw^2\sin^2\beta\cos^2\beta}-{\mt^2\over\sin^2\beta}-
{m_b^2\over\cos^2\beta}\right]\ln\left({\mstop^2\over\mt^2}\right)\,.\cr}
\eqn\onelooplargema
$$
The formula for $\mhl^2$ exhibits the importance of the one-loop
radiative corrections.  The tree-level upper bound, $\mhl\leq\mz$
is significantly modified by terms that are enhanced for large
values of $\mt$ and $\mstop$.  Nevertheless, the numerical value of
the radiative corrections to the squared masses is never
greater than ${\cal O}(\mz^2)$.  Thus, in the limit of $\mha\gg\mz$,
one again finds that $\mha\simeq\mhh\simeq\mhpm$,
up to corrections of ${\cal O}(\mz^2/\mha)$.  One can also show that
$\cos(\beta-\alpha)={\cal O}(\mz^2/\mha^2)$ as before.

\REF\wudka{J.F. Gunion, H.E. Haber, and J. Wudka, {\sl Phys. Rev.}
{\bf D43} (1991) 904.}
The phenomenological implications of these results may be discerned
by reviewing the coupling strengths of the Higgs bosons to
Standard Model particles (gauge bosons, quarks and leptons) in the
two-Higgs doublet model.  The coupling of $\hl$ and $\hh$ to vector
boson pairs or vector-scalar boson final states is proportional to
either $\sin(\beta-\alpha)$ or $\cos(\beta-\alpha)$ as indicated
below.\refmark{\hhg,\hhgsusy}
\vskip 0.2in
\settabs 6 \columns
\+&\us{$\cos(\beta-\alpha)$}&&&\us{$\sin(\beta-\alpha)$}\cr
\vskip 0.1in\+&$\hh\wp\wm$&&&$\hl\wp\wm$\cr
\+&$\hh ZZ$&&&$\hl ZZ$\cr
\+&$Z\ha\hl$&&&$Z\ha\hh$\cr
\+&$W^\pm H^\mp\hl$&&&$W^\pm H^\mp\hh$\cr
\+&$ZW^\pm H^\mp\hl$&&&$ZW^\pm H^\mp\hh$\cr
\+&$\gamma W^\pm H^\mp\hl$&&&$\gamma W^\pm H^\mp\hh$\cr
\vskip 0.2in\noindent
Note in particular that {\it all} vertices
in the theory that contain at least
one vector boson and {\it exactly one} heavy Higgs boson state
($\hh$, $\ha$ or $\hpm$) is proportional to $\cos(\beta-\alpha)$.
This can be understood as a consequence of unitarity sum rules which
must be satisfied by the tree-level amplitudes of the theory.\refmark\wudka\

\def\phm{\phantom{-}}
In models with non-minimal Higgs sectors,
the Higgs couplings to quarks and leptons are model-dependent.
Typically, one imposes constraints on the Higgs-fermion interaction
in order to avoid Higgs mediated tree-level flavor changing neutral
currents.  For example, in the MSSM, one Higgs doublet couples
exclusively to down-type fermions and the second Higgs doublet
couples exclusively to up-type fermions.  In this model, the couplings
of the neutral CP-even Higgs bosons to $f\bar f$ relative to the
Standard Model value are given by (using 3rd family notation)
$$\eqalign{\hl b\bar b:\qquad -{\sin\alpha\over\cos\beta}=\ &\sin(\beta-\alpha)
-\tan\beta\cos(\beta-\alpha)\,,\crr
\hl t\bar t:\qquad \phm{\cos\alpha\over\sin\beta}=\ &\sin(\beta-\alpha)
+\cot\beta\cos(\beta-\alpha)\,,\crr
\hh b\bar b:\qquad \phm{\cos\alpha\over\cos\beta}=\ &\cos(\beta-\alpha)
+\tan\beta\sin(\beta-\alpha)\,,\crr
\hh t\bar t:\qquad \phm{\sin\alpha\over\sin\beta}=\ &\cos(\beta-\alpha)
-\cot\beta\sin(\beta-\alpha)\,.\cr}\eqn\hffcoup$$
In contrast to the Higgs couplings to vector bosons,
none of the couplings in eq.~\hffcoup\ vanish when $\cos(\beta-
\alpha)=0$.  This is a model-independent feature
of the Higgs couplings to fermions.  One finds a similar behavior for
the Higgs self-couplings.  Namely,  the various Higgs self-couplings are
model-dependent since they depend on the form of the scalar
potential.  Nevertheless, one can show that the Higgs
self-couplings do not vanish when $\cos(\beta-\alpha)=0$.

The significance of $\cos(\beta-\alpha)= 0$ is now evident: in this limit,
%In the MSSM, Higgs phases have been chosen such that $0\beta\pi/2$
%and $-\pi/2\leq\alpha\leq 0$.  Thus $\cos(\beta-\alpha)\simeq 0$
%implies that $\alpha\simeq\beta-\pi/2$.  Inserting this result in
%the couplings above, one easily checks that the couplings of $\hl$ to the
couplings of $\hl$ to
gauge boson pairs and fermion pairs are identical to the couplings of
the Higgs boson in the minimal Standard Model.\foot{Likewise, the
$\hl\hl\hl$ and $\hl\hl\hl\hl$ couplings also reduce to their Standard Model
values when $\cos(\beta-\alpha)=0$.}   More precisely, in
the limit of $\mha\gg\mz$, the effects of the heavy Higgs states
($\hpm$, $\hh$ and $\ha$) decouple, and the low-energy effective
scalar sector is indistinguishable from that of the minimal
Standard Model.

\REF\sugra{See, \eg, M. Carena, M. Olechowski, S. Pokorski, and
C.E.M. Wagner, {\sl Nucl. Phys.} {\bf B419} (1994) 213;  G.L. Kane,
C. Kolda, L. Roszkowski and  J.D. Wells, {\sl Phys. Rev.}
{\bf D49} (1994) 6173; {\bf D50} (1994) 3498;  P. Nath and R. Arnowitt,
CERN-TH.7227/94 (1994) [hep-ph 9406403];  W. de Boer, R. Ehret,
W. Oberschulte, D.I. Kazakov, IEKP-KA/94-05 [hep-ph 9405342], to
be published in {\sl Z. Phys. C}.}
In the MSSM, the decoupling regime ($\mha\gg\mz$) sets in rather
quickly once $\mha$ is taken above $\mz$.  Although $\mha$ is a
free parameter of the MSSM, its origin is intimately connected to the
scale of supersymmetry breaking.  From eq.~\monetwoeq, we see that $\mha^2$
is proportional to the soft-supersymmetry breaking parameter
$m_{12}^2$.   Generically, one would expect that $\mha$ is of the same order
as a typical soft-supersymmetry-breaking mass
parameter.
%If the mass scale of typical supersymmetry-breaking
%parameters is somewhat higher than $\mz$, it seems reasonable to
%expect that $\mha$ will be similar in magnitude to other
%supersymmetric particle masses.
In supergravity based models, $m_{12}^2\equiv B\mu$, where $B$ is
a soft-supersymmetry-breaking parameter.  Models  of this type with
universal soft-supersymmetry-breaking scalar masses at the Planck
scale tend to yield values of
$\mha$  that typically lie above $\mz$.\refmark\sugra\
In such approaches, one would expect the couplings of
$\hl$ to be nearly identical to those of the Standard Model Higgs boson.

\bigskip
\REF\thomas{H.E. Haber and S. Thomas, SCIPP preprint in preparation.}
\REF\nir{H.E. Haber and Y. Nir, {\sl Nucl. Phys.} {\bf B335}
(1990) 363.}
\chapter{Decoupling Properties of the Two-Higgs Doublet Model\refmark\thomas}
\medskip

The decoupling properties of the MSSM Higgs sector are not special to
supersymmetry.  Rather, they are a generic feature of non-minimal
Higgs sectors.\refmark\nir\  In this section, I demonstrate this
assertion in the case of the most general CP-conserving two-Higgs doublet
model.
Let $\Phi_1$ and
$\Phi_2$ denote two complex $Y=1$, SU(2)$\ls{L}$ doublet scalar
fields.\foot{In terms of the $Y=\pm1$ fields of the previous section,
$H_1^i=\epsilon_{ij}{\Phi_1^j}^\star$ and $H_2=\Phi_2$.}
The most general gauge invariant scalar potential is given by\foot{%
In most discussions of two-Higgs-doublet models, the terms proportional
to $\lambda_6$ and $\lambda_7$ are absent at tree-level.  This can be
achieved by
imposing a discrete symmetry $\Phi_1\rta -\Phi_1$ on the model.  Such a
symmetry would also require $m_{12}=0$ unless we allow a
soft violation of this discrete symmetry by dimension-two terms.
This latter requirement is sufficient to guarantee the absence of
Higgs-mediated tree-level flavor changing neutral currents.}
$$\eqalign{
\calv&=m_{11}^2\Phi_1^\dagger\Phi_1+m_{22}^2\Phi_2^\dagger\Phi_2
-[m_{12}^2\Phi_1^\dagger\Phi_2+{\rm h.c.}]\crrr
&\quad +\half\lambda_1(\Phi_1^\dagger\Phi_1)^2
+\half\lambda_2(\Phi_2^\dagger\Phi_2)^2
+\lambda_3(\Phi_1^\dagger\Phi_1)(\Phi_2^\dagger\Phi_2)
+\lambda_4(\Phi_1^\dagger\Phi_2)(\Phi_2^\dagger\Phi_1)\crrr
&\quad +\left\{\half\lambda_5(\Phi_1^\dagger\Phi_2)^2
+\big[\lambda_6(\Phi_1^\dagger\Phi_1)
+\lambda_7(\Phi_2^\dagger\Phi_2)\big]
\Phi_1^\dagger\Phi_2+{\rm h.c.}\right\}\,.\cr}\eqn\pot$$\vskip5pt\noindent
In principle, $m_{12}^2$, $\lambda_5$,
$\lambda_6$ and $\lambda_7$ can be complex.  However, I shall
ignore the possibility of CP-violating effects in the Higgs sector
by choosing all coefficients in eq.~\pot\ to be real.
The scalar fields will
develop non-zero vacuum expectation values if the mass matrix
$m_{ij}^2$ has at least one negative eigenvalue. Imposing CP invariance
and U(1)$\ls{\rm EM}$ gauge symmetry, the minimum of the potential is
\vskip3pt
$$
\langle \Phi_1 \rangle={1\over\sqrt{2}}
\pmatrix{0\cr v_1\cr}, \qquad \langle \Phi_2\rangle=
{1\over\sqrt{2}}\pmatrix{0\cr v_2\cr}\,,\eqn\potmin$$
\vskip3pt\noindent
where the $v_i$ can be chosen to be real and positive.
It is convenient to introduce the following notation:
$$
v^2\equiv v_1^2+v_2^2={4\mw^2\over g^2}\,,
\qquad\qquad t_\beta\equiv\tanb\equiv{v_2\over v_1}\,.\eqn\tanbdef$$
%Of the original eight scalar degrees of freedom, three Goldstone
%bosons ($G^\pm$ and $G^0$)
%are absorbed (``eaten'') by the $W^\pm$ and $Z$.  The remaining
%five physical Higgs particles are: two CP-even scalars ($\hl$ and
%$\hh$, with $\mhl\leq \mhh$), one CP-odd scalar ($\ha$) and a charged
%Higgs pair ($\hpm$).
The mass parameters $m_{11}^2$ and $m_{22}^2$ can be
eliminated by minimizing the scalar potential.  The resulting
squared masses for the CP-odd and charged Higgs states are
\vskip4pt
$$\eqalignno{%
\mha^2 &={m_{12}^2\over \sb\cb}-\half
v^2\big(2\lambda_5+\lambda_6 t_\beta^{-1}+\lambda_7t_\beta\big)\,,
&\eqnalign\massha \cr\crr
m_{H^{\pm}}^2 &=m_{A^0}^2+\half v^2(\lambda_5-\lambda_4)\,,
&\eqnalign\mamthree\cr}$$
\vskip5pt\noindent
where $s_\beta\equiv\sin\beta$ and $c_\beta\equiv\cos\beta$.
The two CP-even Higgs states mix according to the following squared mass
matrix:
$$
\calm^2 =m_{A^0}^2 \left(\matrix{\sb^2&-\sb\cb\cr
-\sb\cb&\cb^2}\right)+v^2\left(\matrix{\calm^2_{11}&\calm^2_{12}\cr
\calm^2_{12}&\calm^2_{22}\cr}\right)\,,\eqn\massmhh$$
where
$$\left(\matrix{\calm^2_{11}&\calm^2_{12}\cr
\calm^2_{12}&\calm^2_{22}\cr}\right)\equiv
\left( \matrix{\lambda_1\cb^2+2\lambda_6\sb\cb+\lambda_5\sb^2
&(\lambda_3+\lambda_4)\sb\cb+\lambda_6
\cb^2+\lambda_7\sb^2\crr
(\lambda_3+\lambda_4)\sb\cb+\lambda_6
\cb^2+\lambda_7\sb^2&
\lambda_2\sb^2+2\lambda_7\sb\cb+\lambda_5\cb^2}\right)
\eqn\calmdef$$
%The physical mass eigenstates are
%$$\eqalign{%
%\hh &=(\sqrt2\, {\rm Re\,}\Phi_1^0-v_1)\cos\alpha+
%(\sqrt2\,{\rm Re\,}\Phi_2^0-v_2)\sin\alpha\,,\crr
%\hl &=-(\sqrt2\,{\rm Re\,}\Phi_1^0-v_1)\sin\alpha+
%(\sqrt2\,{\rm Re\,}\Phi_2^0-v_2)\cos\alpha\,.\cr}
%\eqn\scalareigenstates$$
%The corresponding masses are
%$$
% m^2_{\hh,\hl}=\half\left[{\cal M}_{11}^2+{\cal M}_{22}^2
%\pm \sqrt{({\cal M}_{11}^2-{\cal M}_{22}^2)^2 +4({\cal M}_{12}^2)^2}
%\ \right]\,,
%\eqn\higgsmasses$$
%and the mixing angle $\alpha$ is obtained from
%$$\eqalign{%
%\sin 2\alpha &={2{\cal M}_{12}^2\over
%\sqrt{({\cal M}_{11}^2-{\cal M}_{22}^2)^2 +4({\cal M}_{12}^2)^2}}\ ,\crrr
%\cos 2\alpha &={{\cal M}_{11}^2-{\cal M}_{22}^2\over
%\sqrt{({\cal M}_{11}^2-{\cal M}_{22}^2)^2 +4({\cal M}_{12}^2)^2}}\ .\cr}
%\eqn\alphadefff$$
It is convenient to define four squared mass combinations:
$$\eqalign{\mlsq\equiv&\ \calm^2_{11}\cos^2\beta+\calm^2_{22}\sin^2\beta
+\calm^2_{12}\sin2\beta\,,\crr
\mdsq\equiv&\ \left(\calm^2_{11}\calm^2_{22}-\calm^4_{12}\right)^{1/2}\,,\crr
\mtsq\equiv&\ \calm^2_{11}+\calm^2_{22}\,,\crr
\mssq\equiv&\ \mha^2+\mtsq\,.\cr}
\eqn\massdefs$$
In terms of these quantities,
$$ m^2_{\hh,\hl}=\half\left[\mssq\pm\sqrt{m\ls{S}^4-4\mha^2\mlsq
-4m\ls{D}^4}\,\right]\,,\eqn\cpevenhiggsmasses$$
and
$$\cos^2(\beta-\alpha)= {\mlsq-\mhl^2\over\mhh^2-\mhl^2}\,.
%\half\left[1+{2\mlsq-\mssq\over
%\mhh^2-\mhl^2}\right]
\eqn\cosbmasq$$

Suppose that all the Higgs self-coupling constants $\lambda_i$ are
held fixed such that $\lambda_i\lsim1$, while taking
$\mha^2\gg\lambda_iv^2$.  Then the $\calm^2_{ij}\sim{\cal O}(v^2)$, and
it follows that:
$$\mhl\simeq m\ls{L}\,,\qquad\qquad \mhh\simeq\mha\simeq\mhpm\,,
\eqn\approxmasses$$
and
$$\cos^2(\beta-\alpha)\simeq\, {\mlsq(\mtsq-\mlsq)-m\ls{D}^4\over\mha^4}\,.
\eqn\approxcosbmasq$$
Comparing these results with those of eq.~\largema, one sees that the
MSSM results are simply a special case of the more general
two-Higgs doublet model results just obtained.

The limit $\mha^2\gg\lambda_iv^2$ (subject to $\lambda_i\lsim 1$) will be
called the {\it decoupling limit} of the model.  From eq.~\massha, it
follows that $m_{12}^2\gg\lambda_iv^2$ (assuming that neither $t_\beta$
nor $t^{-1}_\beta$ is close to 0).  This condition clearly
depends on the original choice of scalar field basis $\Phi_1$ and
$\Phi_2$.  For example, I can diagonalize the squared mass terms of
the scalar potential [eq.~\pot] thereby setting $m_{12}=0$.  In the
decoupling limit in the new basis, eq.~\massha\ would then imply that
$\beta$ must be near 0 or near $\pi/2$.  A basis independent
characterization of the decoupling limit is simple to formulate.
Starting from the scalar potential in an arbitrary basis,
form the matrix $m_{ij}^2$.
%$\pmatrix{m_{11}^2&-m_{12}^2\cr -m_{12}^2&m_{22}^2 \cr}$.
Denote the eigenvalues of this matrix by $m_a^2$ and $m_b^2$
respectively; note that the eigenvalues are real but can be of either
sign.  By convention, I shall take $|m_a^2|\leq|m_b^2|$.  Then, the
decoupling limit corresponds to $m_a^2<0$, $m_b^2>0$ such that
$m_b^2\gg |m_a^2|, v$ (with $\lambda_i\lsim1$).

We are now ready to consider the questions posed in section 1.
Let us assume that $\mhl\ll\mhh,\mha,\mhpm$.  From
eq.~\cpevenhiggsmasses, it follows that
$$0<\mha^2\mlsq+m\ls{D}^4\ll m\ls{S}^4\,.\eqn\massinequality$$
%We are interested in whether the properties of $\hl$ can be
%distinguished from the Standard Model Higgs boson.
Eq.~\approxcosbmasq\ implies
that in the decoupling limit, $\cos(\beta-\alpha)=
{\cal O}(\mz^2/\mha^2)$, which means that the $\hl$ couplings to
Standard Model particles match precisely those of the Standard Model
Higgs boson.  However, in the following, I shall {\it not}
impose the decoupling limit.
Rather, I shall only require eq.~\massinequality, which
reflects my basic assumption that one scalar state, $\hl$,  is
lighter than the other Higgs bosons.
Eq.~\massinequality\ is satisfied in one of two cases:\item{(i)}$\mz^2$,
$\mlsq$, $m\ls{D}^4/\mha^2\ll \mha^2$, $\mssq$.
That is, each term on the left-hand side of eq.~\massinequality\ is
separately smaller in magnitude than $m\ls{S}^4$, or
\item{(ii)} $\mha^2\mlsq$ and $m\ls{D}^4$ are both of ${\cal O}(m\ls{S}^4)$,
but due to cancelation of the leading behavior of each term, the sum
satisfies eq.~\massinequality.

\vskip 0.1in
\noindent In case (i), one finds that
$$\mhl^2\simeq {\mha^2\mlsq\over\mssq}+{m\ls{D}^4\over\mssq}
+{\mha^4m\ls{L}^4\over m\ls{S}^6}+{\cal O}\left({m\ls{L}^4\over
m\ls{S}^4}\right)\,,\eqn\mhlcasei$$
and $\mhh^2\sim{\cal O}(\mssq)$.  In the same approximation,
$$\cos^2(\beta-\alpha)\simeq{\mlsq\over\mssq}\left(1-
{\mha^2\over\mssq}\right)+{1\over m\ls{S}^4} \left[m\ls{L}^4\left(
{2\mha^2\over\mssq}-{3\mha^4\over m\ls{S}^4}\right)
-m\ls{D}^4\right]\,.
\eqn\cbmacasei$$
The behavior of $\cos(\beta-\alpha)$ depends crucially on how close
$\mha^2/\mssq$ is to 1.  If $\mtsq\ll\mssq$ [see eq.~\massdefs],
we recover the results of the decoupling limit [eqs.~\approxmasses\
and \approxcosbmasq].  On the other
hand, if $\mtsq\sim{\cal O}(\mssq)$, then eq.~\cbmacasei\ implies
that $\cos(\beta-\alpha)\sim{\cal O}(\mz/\mha)$.  This is a particular
region of the parameter space where some of the $\lambda_i$ are
substantially larger than 1, and yet the $\hl$ couplings do not
significantly deviate from those of the Standard Model.
Nevertheless, the onset of decoupling is slower than the
$\cos(\beta-\alpha)\sim{\cal O}(\mz^2/\mha^2)$ behavior found in the
decoupling regime.   In order to find a parameter regime which
exhibits complete non-decoupling, one must consider case (ii) above.
In this case, despite the fact that $\mhl\ll\mhh,\mha,\mhpm$, one
nevertheless finds that $\cos(\beta-\alpha)\sim{\cal O}(1)$, which implies
that the couplings of $\hl$ deviate significantly from those of the
Standard Model Higgs boson.  Although it might appear that case (ii)
requires an unnatural cancelation, it is easy to construct a simple
model of non-decoupling.  Consider a model where:
$m_{12}^2=\lambda_6=\lambda_7=0$, $\lambda_3+\lambda_4+\lambda_5=0$,
$\lambda_2\lsim{\cal O}(1)$, and $\lambda_1, \lambda_3, -\lambda_5\gg1$.
This model yields: $\mhl^2=\lambda_2 v^2 s_\beta^2$,
$\mhh^2=\lambda_1 v^2
c_\beta^2$, $\mha^2=-\lambda_5 v^2$, $\mhpm^2=\half\lambda_3 v^2$,
and $\cos^2(\beta-\alpha)=c_\beta^2$.  Note that in this model,
the heavy Higgs states are not approximately degenerate (as required in the
decoupling limit).

\bigskip
\chapter{Phenomenological Challenges of the Decoupling Limit}
\medskip

We have seen that in the decoupling limit, the couplings of $\hl$ to
Standard Model gauge bosons and fermions approach those of the
Standard Model Higgs boson.  Suppose that a future experiment has
already discovered and studied the properties of $\hl$.
What are the requirements of experiments at future colliders
for proving the existence or non-existence of a non-minimal Higgs sector?
To be specific, let us assume in this section that $\hl$ has been
discovered with couplings approximating those of the Standard Model
Higgs boson and $\mha>250$~GeV.

\REF\higgstt{D. Dicus, A. Stange and S. Willenbrock, {\sl Phys. Lett.}
{\bf B333} (1994) 126.}
\REF\xsecs{For a recent review, see
J.F. Gunion, in {\it Perspectives on Higgs Physics},
edited by G.L. Kane (World Scientific, Singapore, 1993) pp.~179--222.}
\REF\vega{J. Dai, J.F. Gunion and R. Vega, {\sl Phys. Rev. Lett.}
{\bf 71} (1993) 2699; UCD-94-7 (1994) [hep-ph 9403362].}
At the LHC, the rate for $gg\rta \ha$, $\hh$ and $gb\rta\hm t$ provides
a substantial number of produced Higgs bosons per year (assuming
that the heavy Higgs masses are not too large).\refmark\xsecs\
Unfortunately, there may not be a viable final state signature.  For
example, since $\cos^2(\beta-\alpha)\ll 1$, the branching ratio
of $\hh\rta ZZ$ is significantly suppressed, so that the gold-plated
Standard Model Higgs signature is simply absent.  At present, there
is no known proven technique for detecting $\ha$, $\hh$ and $\hpm$ signals
at the LHC in the decoupling regime of parameter space.  An
attempt to isolate a Higgs signal in $t\bar t$ final states has been
discussed in ref.~\higgstt.  Another method consists of a search for Higgs
signals in $t\bar tt\bar t$, $t\bar t b\bar b$, and $b\bar bb\bar b$
final states.\refmark\vega\
These can arise from $gg\rta Q\bar Q^\prime(\hh$, $\ha$ or
$\hpm$), where $Q$ is a heavy quark ($b$ or $t$), and the Higgs boson
subsequently decays into a heavy quark pair.  As noted
below eq.~\hffcoup, even in the decoupling
limit, the couplings of $\hh$, $\ha$ and $\hpm$ to heavy quark pairs
are not suppressed.  Whether such signals can be extracted from the
substantial QCD backgrounds (very efficient $b$-tagging is one of the
many requirements for such a search) remains to be seen.

\REF\janot{P. Janot, in
{\it Physics and Experiments with Linear $\epem$ Colliders},
Workshop Proceedings, Waikoloa, Hawaii, 26--30 April, 1993,
edited by F.A. Harris, S.L. Olsen, S. Pakvasa and X. Tata
(World Scientific, Singapore, 1993) pp.~192--217.}
\REF\nlcsearch{A. Djouadi, J. Kalinowski and P.M. Zerwas, {\sl Z. Phys.}
{\bf C57} (1993) 569.}
Let us now turn to $e^+e^-$ colliders.  First, consider the process
$e^+e^-\rta\hl\ha$ (via $s$-channel $Z$-exchange).  Since the $Z\hl\ha$
coupling is proportional to $\cos(\beta-\alpha)$, the
production rate is suppressed in the decoupling regime.  For example,
in the MSSM, if $\mha>\mhl$, then LEP-II will not
possess sufficient energy and/or luminosity to directly produce the
$\ha$.\refmark\janot\  Of course, with sufficient energy, one can directly
pair-produce the heavy Higgs states via $e^+e^-\rta H^+H^-$ and
$e^+e^-\rta\hh\ha$ without a rate suppression.
At the NLC, with $\sqrt{s}=500$~GeV and 10~ fb$^{-1}$ of
data, it has been shown\refmark{\janot,\nlcsearch}
that no direct signals for $\ha$, $\hh$, and
$\hpm$ can be seen if $\mha\gsim 230$~GeV.  Although this result was
obtained in the MSSM, it also applies to the decoupling regime of
more general Higgs sectors.  These results seems to imply the
following rather general conclusion: {\it evidence for the non-minimal
Higgs sector at the NLC requires a machine with energy
$\sqrt{s}>2\mha$, sufficient to pair-produce heavy Higgs states}.

\REF\djouadi{A. Djouadi, J. Kalinowski and P.M. Zerwas,
{\sl Mod. Phys. Lett.} {\bf A7} (1992) 1765; {\sl Z. Phys.} {\bf C54}
(1992) 255.}
Can this conclusion be avoided?  There are two possible methods.
\pointbegin
Produce one heavy Higgs state in association with
light states.\foot{In this context, light states refer to all Standard Model
fermions and bosons, with masses of order $\mz$ or less.  Thus,
the gauge bosons, $\hl$, and even the top-quark will be considered
among the light states!}.
\point
Make precision measurements of $\hl$ couplings to
Standard Model particles in order to detect a deviation from the
Standard Model expectations.

\noindent
First, consider production mechanisms which result in a singly produced
heavy Higgs state.  It was noted in section 2 that whenever a single
heavy Higgs state couples directly to a gauge boson plus other
particles, the coupling is suppressed by $\cos(\beta-\alpha)$.
To avoid this suppression, one must couple the heavy Higgs state to
either fermions or scalars.  For example, consider $e^+e^-\rta Q\bar
Q^\prime(\hh,\ha$, or $\hpm)$, where $Q=b$ or $t$.  The production rates
have been computed by Djouadi \etal\refmark\djouadi\
Unfortunately, the three-body
phase space greatly suppresses the rate once the heavy Higgs state
is more massive than the $Z$.  In particular, for
$\mha>\sqrt{s}/2$, these processes do not provide viable signatures
for the heavy Higgs states.  A similar conclusion is
obtained when the heavy Higgs state couples to light scalars.
Scott Thomas and I have computed the rate for $e^+e^-\rta\hl\hh Z$.
We assumed
that the dominant contribution arises in the $s$-channel $Z$-exchange,
where the virtual $Z^\ast$ decays via $Z^\ast\rta Z{\hl}^\ast\rta Z\hl\hh$.
In the limit where $\mhh\gg\mhl,\mz$, we obtained\refmark\thomas\
$$\eqalign{
{\sigma(e^+e^-\rta\hl\hh Z)\over\sigma(e^+e^-\rta\hl Z)}\simeq
{g^2_{\hh\hl\hl}\over 32\pi^2s^3\mhh^2}\biggl\{(s-\mhh^2)&\left[(s-\mhh^2)^2
+12s\mhh^2\right]\cr
-6\mhh^2& s(s+\mhh^2)\ln\left({s\over\mhh^2}\right)\biggr\}\,,\cr}
\eqn\threeh$$
where the $\hh\hl\hl$ coupling in the decoupling limit
[\ie, when $\cos(\beta-\alpha)=0$] is given by
$$g_{\hh\hl\hl}={-3ig\mz\over 4\cos\theta\ls{W}}\sin 4\beta\,.\eqn\hhh$$
This rate [eq.~\threeh] is also too small for a viable Higgs signal.

\REF\burke{M.D. Hildreth, T.L. Barklow, and D.L. Burke, {\sl Phys. Rev.}
{\bf D49} (1994) 3441.}
\REF\hildreth{H.E. Haber and M.D. Hildreth, in preparation.}
Second, consider precision measurements of $\hl$ branching ratios at
the NLC.  In a Monte Carlo analysis, Hildreth \etal\refmark\burke\
evaluated the
anticipated accuracy of $\hl$ branching ratio measurements at the NLC,
assuming $\sqrt{s}=500$~GeV and a data set of 50~fb$^{-1}$.   For
example, taking $\mhl=120$~GeV, they computed an extrapolated error
of $\pm 7\%$ for the $1$-$\sigma$ uncertainty in $BR(\hl\rta b\bar b)$
and $\pm 14\%$ for $BR(\hl\rta\tau^+\tau^-)$.  Other channels yielded
substantially larger uncertainties.  To see whether these are
significant measurements, one can compare these results with the
theoretical expectations of the MSSM as a function of $\mha$ and
$\tanb$.  Hildreth and I have found\refmark\hildreth\ that deviations in
$BR(\hl\rta b\bar b)$ and $BR(\hl\rta\tau^+\tau^-)$ from
the Standard Model can be about 7\% for values
of $\mha$ as large as 450~GeV and about 15\% for values of $\mha$ as large
as 250~GeV.  (See ref.~\janot\ for a related analysis.)  These results
imply that a precision measurement of $\hl\rta b\bar b$ has the
potential for detecting the existence of the non-minimal Higgs sector
even if the heavier Higgs states cannot be directly detected at the
NLC.  Of course, one will have to push the precision of this
measurement beyond its present expectations, since a 2-$\sigma$
deviation is not compelling evidence for new physics.  Other Higgs
decay channels are not competitive.

\REF\gamgamcol{
H.F. Ginzburg, G.L. Kotkin, V.G. Serbo and V.I. Telnov,
{\sl Nucl. Inst. and Methods} {\bf 205} (1983) 47;
H.F. Ginzburg, G.L. Kotkin, S.L. Panfil,
V.G. Serbo and V.I. Telnov,
{\sl Nucl. Inst. and Methods} {\bf 219} (1984) 5.}
%T.L. Barklow,
%in {\it Research Directions for the Decade}, Proceedings of the 1990
%Summer Study on High Energy Physics, Snowmass, CO, June 25--July
%13, 1990, edited by E.L. Berger (World Scientific, Singapore,
%1992), pp.~440--450.}
\REF\gamgamhiggs{A.I. Vainshtein, M.B. Voloshin, V.I. Zakharov and
M. Shifman, {\sl Yad. Fiz.} {\bf 30} (1979) 1368
[{\sl Sov. J. Nucl. Phys.} {\bf 30} (1979) 711].}

There is one novel approach which could extend the discovery limits
for heavy Higgs bosons at the NLC.  A high energy, high luminosity
photon beam can be produced by the Compton backscatter of an intense
laser beam off a beam of electrons.\refmark\gamgamcol\
This provides a mechanism for
turning the NLC into a high energy, high luminosity $\gamma\gamma$
collider.  All neutral Higgs states couple to $\gamma\gamma$ at one-loop
via loops of charged matter.\refmark\gamgamhiggs\
Since the couplings of the heavy Higgs
states to fermions and scalars are not suppressed in the decoupling
limit, the $\gamma\gamma$
couplings of the heavy Higgs states are also not suppressed relative to
the $\hl\gamma\gamma$ coupling.
%(This fact was also implicitly used
%when we noted that the production rates of the heavy Higgs states via
%gluon-gluon fusion at the LHC were not suppressed.)
Thus, one
can search for the non-minimal Higgs sector at the $\gamma\gamma$ collider by
either measuring the $\hl\gamma\gamma$ coupling with
sufficient precision or by directly producing $\ha$ and/or $\hh$ in
$\gamma\gamma$ fusion.  In the decoupling regime, the
$\hl\gamma\gamma$ coupling approaches the corresponding coupling of
the Standard Model Higgs boson.\foot{The contributions of
supersymmetric loops to the $\hl\gamma\gamma$ amplitude
vanish in the limit of large supersymmetric particle masses.}
As a result, this is not
a viable method for detecting deviations from the Standard
Model.  Thus, one must focus on $\gamma\gamma\rta(\ha,\hh)$.  In
ref.~\ghphoton, Gunion and I showed that parameter regimes exist where
one could extend the heavy Higgs mass reach above $\sqrt{s}/2$.
For example, at a 500~GeV $\gamma\gamma$ collider, a statistically significant
$\ha$ signal in $b\bar b$ and $Z\hl$ final states could be seen
above backgrounds if $\mha<2\mt$.

\bigskip
\chapter{Conclusions}
\medskip

In the most general CP-conserving two Higgs doublet model, a
decoupling limit can be defined in which the lightest Higgs state is
a CP-even neutral Higgs scalar, whose properties approach those of the
Standard Model Higgs boson.  This result is more general, and
applies to non-minimal Higgs sectors that contain the Standard Model
Higgs doublet and respect standard phenomenological constraints
(such as $\mw=\mz\cos\theta_W$ at tree-level).  In the MSSM, the
decoupling limit corresponds to $\mha\gg\mz$ (independent of
$\tanb$).  Moreover, the approach to decoupling is rapid once $\mha$
is larger than $\mz$.  Thus, over a very large range of MSSM
parameter space, the couplings of $\hl$ to
gauge bosons, quarks and leptons are nearly identical
to the couplings of the Standard Model Higgs boson.
%Moreover, large $\mha$ corresponds to a large soft-supersymmetry-%
%breaking mass term $m_{12}^2$, which is expected to be large in some
%approaches.

If the $\hl$ is discovered with properties approximating those of the
Standard Model Higgs boson, then the discovery of the non-minimal Higgs
sector will be difficult.  At the LHC, $\ha$, $\hh$ and $\hpm$
production rates via gluon-gluon fusion are not suppressed.  However,
isolating signals of the heavy Higgs states above background presents
a formidable challenge.  At the NLC, if $\sqrt{s}>2\mha$, then pair
production of $\hp\hm$ and $\hh\ha$ is easily detected.   However,
below pair-production threshold, detection of the non-minimal Higgs
sector is problematical.  For example, the
cross sections for single heavy Higgs boson production (in association
with light particles) are too
small to be observed.  However,
experiments at the $\gamma\gamma$ collider
may extend the NLC discovery limits of the heavy Higgs states (via
$\gamma\gamma$ fusion to $\hh$ or $\ha$).  Precision measurements of
$\hl\rta b\bar b$ could provide additional evidence for a non-minimal Higgs
sector.  However, current experimental expectations at the NLC
predict only a 2-$\sigma$ deviation from Standard Model expectations
if $\mha$ is just beyond the kinematic limit of direct NLC detection.

\REF\valle{A.S. Joshipura and J.W.F. Valle, {\sl Nucl. Phys.}
{\bf B397} (1993) 105.}
\REF\griest{K. Griest and H.E. Haber, {\sl Phys. Rev.}
{\bf D37} (1988) 719.}
If $\hl$ is discovered with distinguishable properties relative to
Standard Model predictions, then there are three possibilities.
The simplest possibility is that other scalar states of the non-%
minimal Higgs sector lie close in mass to $\hl$ and will be
discovered via direct production soon after the discovery of $\hl$.
If the other Higgs states are not
discovered (and if appropriately strong mass limits are obtained),
then two alternative scenarios emerge.   If the Higgs parameters
lie in the non-decoupling regime (\ie, there exists
at least one or more large Higgs
self-couplings), then the Higgs sector is probably strongly coupled, which
suggests the existence of a rich electroweak
symmetry breaking sector at an energy scale near 1~TeV.
On the other hand, if the Higgs sector is weakly coupled, then
there must exist new
non-Standard Model decay channels accessible in $\hl$ decay.  For
example, in Majoron models, $\hl\rta JJ$ (where $J$ is the Majoron)
provides an invisible decay mode for $\hl$,\refmark\valle\
which would cause
deviations from Standard Model expectations
for $\hl$ branching ratios to gauge bosons, quarks and
leptons.  A similar possibility exists in some supersymmetric models
where the decay $\hl\rta\widetilde\chi^0_1\widetilde\chi^0_1$
(where $\widetilde\chi^0_1$ is the
lightest neutralino) can be significant,\refmark\griest\
leading again to deviations in the $\hl$ branching ratios to Standard
Model final states.

Once the first evidence for the Higgs boson is established, it will be
crucial to ascertain the underlying dynamics of the electroweak
symmetry breaking sector.  If the data reveals that the
Higgs sector is non-minimal, then we will have an
important clue to the structure of the electroweak symmetry
breaking sector.
However, one must be prepared for the more
pessimistic scenario---the discovery of a Higgs boson whose
properties are not experimentally distinguishable from the  Higgs
boson of the minimal Standard Model.  This scenario presents a formidable
challenge for future collider experiments in their attempt to probe
the physics of the electroweak symmetry breaking sector and the
nature of TeV-scale physics beyond the Standard Model.

\vskip\chapterskip
\centerline{\bf Acknowledgments}
\vskip .1in
This paper describes work performed in a number of separate
collaborations with Jack Gunion, Mike Hildreth,
Yosef Nir, and Scott Thomas.  I am especially
grateful for their insights and diligence.
In addition, I am pleased to acknowledge George
Pocsik and his colleagues at E\"otv\"os University and Stefan
Pokorski and his colleagues at the University of Warsaw for
their warm hospitality and their labors in providing such a
stimulating environment during their workshops.
This work was supported in part by the Department of Energy.

\refout
\bye